\documentclass[sigconf,authorversion,nonacm]{acmart}
\AtBeginDocument{%
  \providecommand\BibTeX{{%
    \normalfont B\kern-0.5em{\scshape i\kern-0.25em b}\kern-0.8em\TeX}}}

\begin{document}

\title{Innovative Digital Storytelling with AIGC: Exploration and Discussion of Recent Advances}


\author{Rongzhang Gu}
\affiliation{%
  \institution{Shanghai AI Laboratory}
} 
\email{gurongzhang@pjlab.org.cn}

\author{Hui Li}
\authornote{Both authors contributed to this research during internships at Shanghai AI Laboratory.}
\affiliation{%
  \institution{Shanghai Theatre Academy}
  \institution{Shanghai AI Laboratory}
} 
\email{lihui1@pjlab.org.cn}

\author{Changyue Su}
\authornotemark[1]
\affiliation{%
  \institution{New York University}
  \institution{Shanghai AI Laboratory}
} 
\email{tinasucy6@gmail.com}

\author{Wayne Wu}
\affiliation{%
  \institution{Shanghai AI Laboratory}
} 
\email{wuwenyan0503@gmail.com}

\renewcommand{\shortauthors}{Gu et al.}


\begin{abstract}
  Digital storytelling, as an art form, has struggled with cost-quality balance. The emergence of AI-generated Content (AIGC) is considered as a potential solution for efficient digital storytelling production. However, the specific form, effects, and impacts of this fusion remain unclear, leaving the boundaries of AIGC combined with storytelling undefined. This work explores the current integration state of AIGC and digital storytelling, investigates the artistic value of their fusion in a sample project, and addresses common issues through interviews. Through our study, we conclude that AIGC, while proficient in image creation, voiceover production, and music composition, falls short of replacing humans due to the irreplaceable elements of human creativity and aesthetic sensibilities at present, especially in complex character animations, facial expressions, and sound effects\footnote{Project page: \hyperlink{https://lsgm-demo.github.io/Leveraging-recent-advances-of-foundation-models-for-story-telling/}{https://lsgm-demo.github.io/Leveraging-recent-advances-of-foundation-models-for-story-telling/}}. The research objective is to increase public awareness of the current state, limitations, and challenges arising from combining AIGC and digital storytelling.
\end{abstract}








\maketitle

\section{Introduction}

Digital storytelling plays a vital role in the contemporary multimedia society, permeating various facets of today’s Internet, offering substantial values across different objectives, including concept explanation, personal experience reflection, and political argument. As articulated by Joe Lambert~\cite{joe2006digital}, the core of digital stories is ``bringing narratives to life'', which properly demonstrates the fundamental elements of its creation process. Digital storytelling conventionally leverages a sophisticated amalgamation of multimedia content to craft attractive, immersive, and interactive experiences, involving several pivotal components such as narratives, storyboards, animation, video, and audio~\cite{robin2008digital}.

\begin{figure*}[htb]
    \centering
    \includegraphics[width=1.0\textwidth]{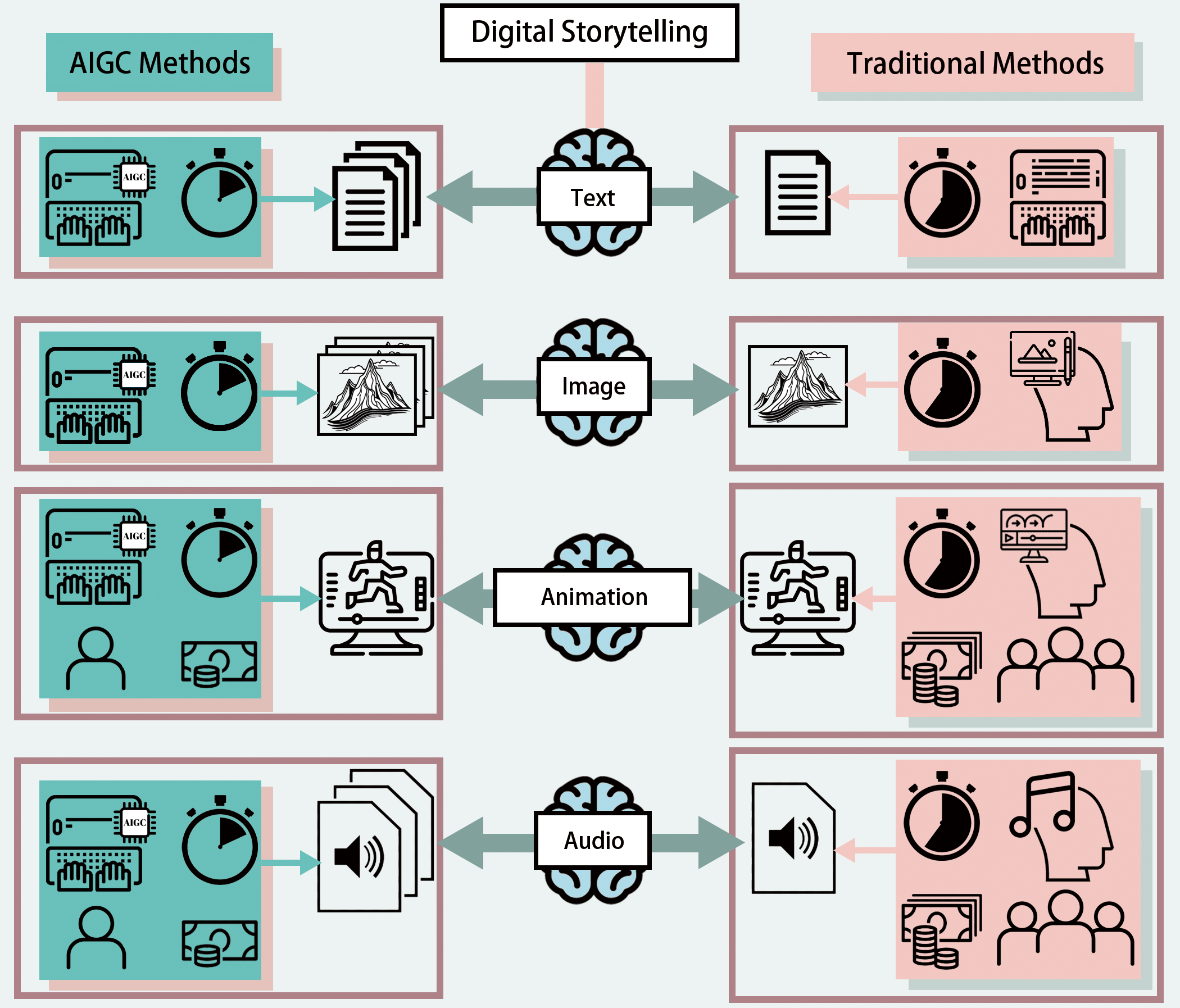}
    \caption{The comparison of the conventional process of computer-assisted digital storytelling creation and the process integrating AIGC. The AIGC approach surpasses the traditional method in terms of output efficiency, time, and resource utilization across four production stages (\textcopyright The author of the paper.)}
    \label{fig:figure1}
    \Description{This figure evaluates input type, time, human costs, and monetary costs when completing digital storytelling modules: Text, Image, Animation, and Audio, using the AIGC Method versus the Traditional Method. Details are provided in supplemental materials.}
\end{figure*}

The production of digital storytelling has been revolutionized by the rapid advancements in computer techniques, including graphics, visualization, and internet platforms. This surge has yielded the development of digital tools, empowering professionals to create increasingly high-quality content. Nevertheless, digital storytelling still requires expertise in various digital art working processes. Recently, the great explosion of AI-generated Content (AIGC) has opened up a compelling avenue for novices to create digital content that satisfies their own ideas with greater ease and efficiency from prompts and sketches, as depicted in Figure~\ref{fig:figure1}. This development augments the capacity of artists in ideation, concept refinement, and content production on an unprecedented scale.

Although AIGC has a range of advantages, the exploration of building comprehensive and efficient AIGC technology chains, while simultaneously ensuring the accurate and meaningful expression of artistic ideas, remains an ongoing endeavor that requires further refinement and investigation. In this study, we do a pioneering experiment called Naked Monkey’s Happy Discovery (N.M.H.D) that involves minimal human intervention, using AIGC as much as possible. Our focus is to preserve nothing but the core idea of the narrative while applying AIGC tools to generate various multimedia elements, such as detailed scripts, character appearances, scene images, animations, audio, and video content. Through this work, we demonstrate a promising technical digital story creation pipeline that heavily relies on AIGC. Additionally, we have provided the website to display further details to the public: \href{https://lsgm-demo.github.io/Leveraging-recent-advances-of-foundation-models-for-story-telling/}{https://lsgm-demo.github.io/Leveraging-recent-advances-of-foundation-models-for-story-telling/}

Despite we have showed a promising path with our AIGC-driven solution, several lingering issues persist, extending from practical to ethical aspects, raising society-wide concerns about the relationships between art and technology. Consequently, we formulate a thorough interview within the realm of creative industries and collect constructive insights regarding the preceding concerns. We summarize the opinions and insights, offering advice about the technological development of AIGC and the evolving landscape of digital storytelling. 

\section{Related Work}

\subsection{Digital Narrative}

Digital storytelling consists of four parts: story script, pictures, audio, and animation~\cite{robin2008digital}. Its flexibility and dynamics incorporate multisensory experiences and utilize cognitive processes for learning~\cite{sadik2008digital}. Animation and Film provide unique paths for narrative expression by creating the sense of a virtual world and incorporating dynamic and visual effects~\cite{chatman1978story}.

\subsection{Technical Development}

Early animation production used manual drawing to create motion trajectories~\cite{lamarre2009anime}. In 1970, computer technologies gradually started to assist animation~\cite{menache2000understanding}. Later, motion capture used trajectory data to create character animation~\cite{sturman1994brief}. Recently, Generative AI models, initially limited to academia, have led to a surge in AI art for the upgrades of some prominent frameworks~\cite{anantrasirichai2022artificial}. The first is the Generative Adversarial Networks, which attracted great attention in the art industry~\cite{goodfellow2014generative}. Then Diffusion Models came out, giving better results and providing multi-modality generation~\cite{croitoru2023diffusion}. Later, more convenient and efficient AI tools emerged, breaching the boundaries between AI and art creation thanks to the development of Large Language Models like GPT~\cite{openai2023gpt4}.

\begin{figure*}[ht]
    \begin{center}
        \includegraphics[width=1.0\textwidth]{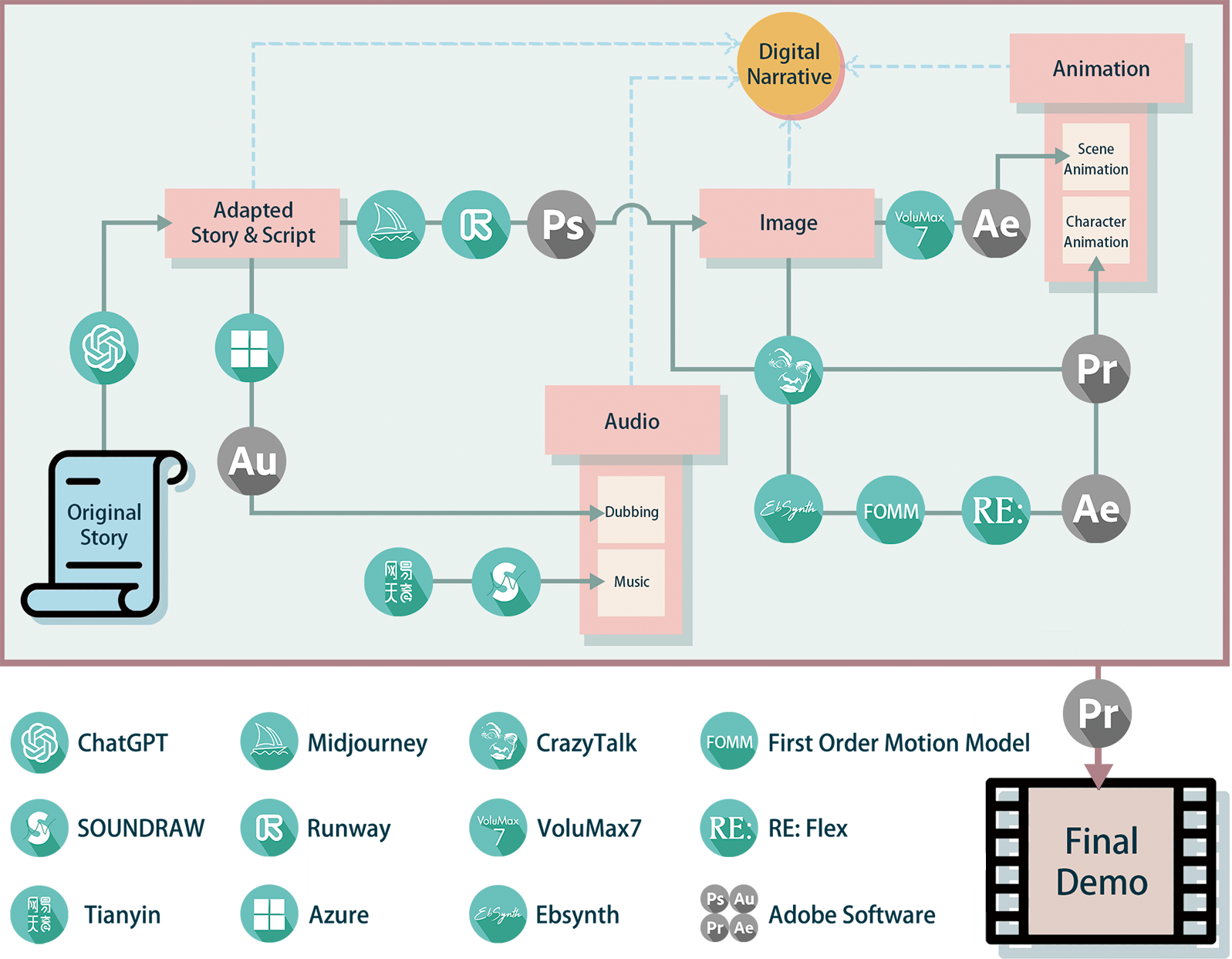}
        \caption{The designed AIGC-driven pipeline for digital storytelling production. We synthesize multimedia intermediate contents with different AIGC tools (green icons) and compose all the results with professional Adobe software (gray icons) for the final demo: \href{https://lsgm-demo.github.io/Leveraging-recent-advances-of-foundation-models-for-story-telling/}{https://lsgm-demo.github.io/Leveraging-recent-advances-of-foundation-models-for-story-telling/} (\textcopyright The author of the paper.)}
        \label{fig:figure2}
        \Description{Flowchart labeled ‘Figure 2’ State Diagram with eight Digital Narrative modules connected by AIGC flow links. The start module is ‘Original Story’. The end module is ‘Final Demo’. Details are provided in supplemental materials.}
    \end{center}
\end{figure*}
\section{AIGC Pipeline}
In this section, we introduce a novel AIGC-based pipeline for creating narrative-centric digital stories using the world’s famous fairy tale “The Emperor's New Clothes” as the adapted example. This fairy tale represents a compelling foundation for AI-driven digital storytelling, owing to a multitude of salient factors. Firstly, its enduring classic status ensures its resonance across diverse cultural and temporal contexts. Secondly, the universally comprehensible themes it encapsulates—namely, hypocrisy, authority, and societal conformity—render it an ideal narrative for engaging with a broad audience. Furthermore, the tale imparts profound moral lessons that have perdured through generations, contributing to its enduring relevance. Its inherent dramatic elements and unexpected plot twists captivate and sustain the interest of viewers, enhancing its suitability for digital storytelling. Importantly, its adaptability for customization to cater to various purposes and target audiences underscores its versatility as a narrative substrate.

As illustrated in Figure~\ref{fig:figure2}, we employ multiple state-of-the-art AIGC models for generating various kinds of digital content. Our process commences with the adapted version of textual narrative generation through the large language model. Then, we use a series of image generation models to create original characters and scenes from text prompts, and we employ diverse AI tools and algorithms to composite scene and character animations based on the generated images. Subsequently, various AI-generated audio elements, such as dubbing and music, are created. Lastly, all the materials are integrated to synthesize the ultimate experimental outcomes.

\subsection{Text Narrative}
\begin{figure}[htbp]
    \centering
    \includegraphics[width=0.3\textwidth]{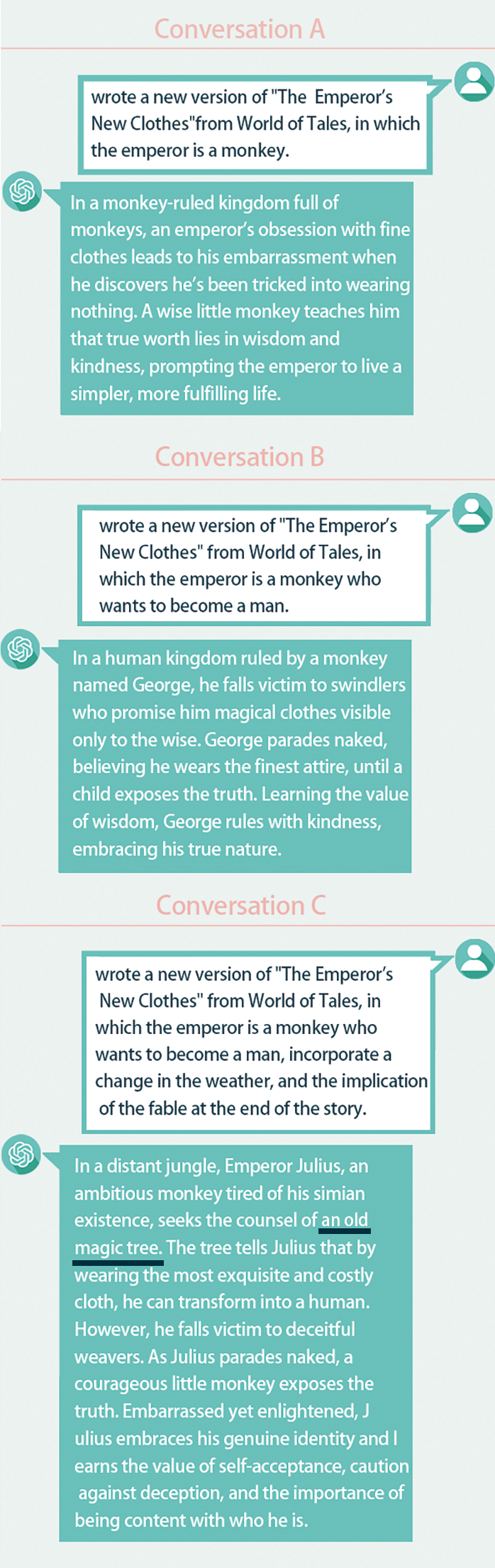}
    \caption{The inquiries and feedback of our three conversations with ChatGPT. We intentionally avoid specifying any predetermined format for the output script or the classification of scenes when inputting text. Instead, we focus on incorporating more restrictions and guidelines to improve the quality of ChatGPT’s feedback (\textcopyright The author of the paper.)}
    \label{fig:figure3}
    \Description{This figure has three columns, demonstrating inquiries and feedback of our three conversations with ChatGPT. Details are provided in supplemental materials.}
\end{figure}

Figure~\ref{fig:figure3} presents the results of our three conversations with ChatGPT. Notably, we deliberately avoided specifying a predetermined format for the output script or the classification of scenes, allowing ChatGPT to autonomously craft the character of the “Old Magic Tree.” This character, which was not originally conceived by humans, plays a crucial role in enhancing the coherence and plausibility of the narrative, adding a distinctive element to the story.

On the one hand, it is important to highlight that the introduction of “Old Magic Tree” exhibits a certain level of randomness (underlined in Figure~\ref{fig:figure3}), which affirms AI’s ability to imagine and generate creative narratives, offering a valuable example to demonstrate AI’s current comprehension and creativity. On the other hand, while ChatGPT excels at summarizing the main points in a text and condensing information~\cite{openai2023gpt4}, the construction of compelling, profound implications or moral lessons within a fable heavily relies on human interpretation, including the understanding of contextual understanding and the skillful interweaving of storytelling elements.

Thus, although we firmly believe that AI’s creative capabilities are well-suited to assist artists in creating more comprehensive and captivating content, human ideas continue to play an indispensable role in the process of text generation in the realm of digital storytelling.

\subsection{Images}
\begin{figure*}[h]
    \centering
    \includegraphics[width=1.0\textwidth]{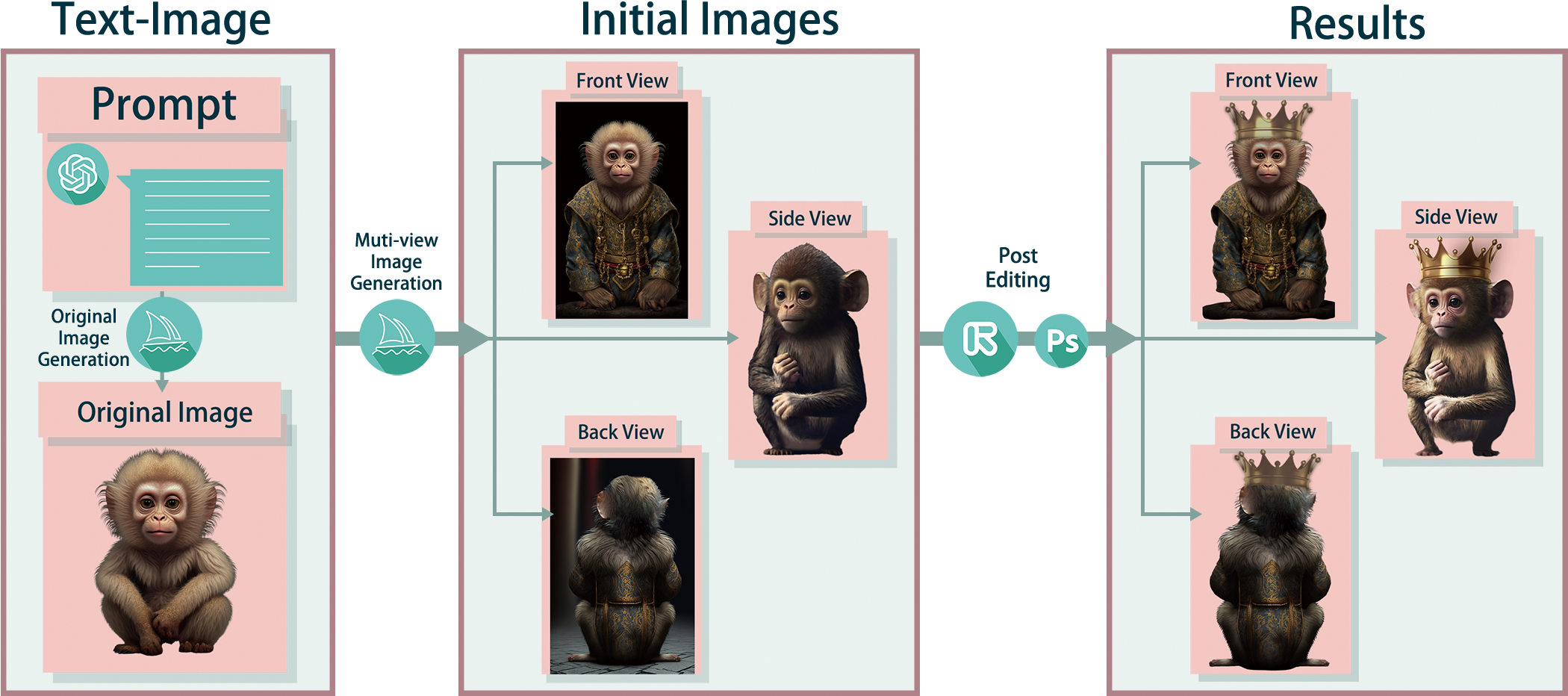}
    \caption{The image generation node is exemplified by creating different perspective images for each character. Initially, the character’s original image is generated by the prompts derived from the story script. Subsequently, the initial multi-view images are generated according to corresponding prompts based on the original image. After modification and refinement processes using AI tools, the final results are completed (\textcopyright The author of the paper.)}
    \label{fig:figure4}
    \Description{The figure demonstrates the image generation pipeline within three steps. The first step is to get the original image from text prompts. The generated images are further transformed into the front, back, and side views. Finally, those different-view images are post-processed with foreground character masks and additional elements like crowns.}
\end{figure*}

The image generation node demonstrated exceptional comprehension and generation skills in capturing the intended content and visual atmosphere. However, human intervention is indispensable to refine the image quality to align with aesthetic preferences. As displayed in Figure~\ref{fig:figure4}, we used Midjourney for raw image generation and Runway for initial adjustments like erasing or replacing suboptimal elements. For a small number of unsatisfactory images, we manually rectified their flaws with Photoshop.

We utilized two Midjourney versions: V4 (pre-V5 release) and V5. While V5 accurately interprets natural language prompts~\cite{midjourney2023userguide}, we applied an AI-transferred step to convert scripts into the specific prompt format for V4 image generation. The general idea is that we informed ChatGPT about Midjourney’s parameters and prompt guidelines first, and then we supplied sample prompts with descriptive phrases from the script, image style, frame ratio, and camera settings as the preconditions for further prompt generation.

Achieving character consistency in diverse scenes and angles was a formidable task in this experimental phase. Initially, we planned to train the character model by establishing the database with stable images that possess the same Midjourney seed value. However, this method proved difficult due to the open-ended nature of our prompt. Therefore, we manually refined the collection of the most prompt-related images in Runway and Photoshop, addressing concerns like extra fingers and missing elements, while ensuring the original character’s features remained intact.

During this node, we observed significant improvements in prompt understanding and image quality when using AIGC for image generation, surpassing its previous performance. However, there is still considerable untapped potential for its further development in generation consistency and capturing intricate details.

\subsection{ Scene and Character Animation}
\begin{figure*}[h]
    \centering
    \includegraphics[width=1.0\textwidth]{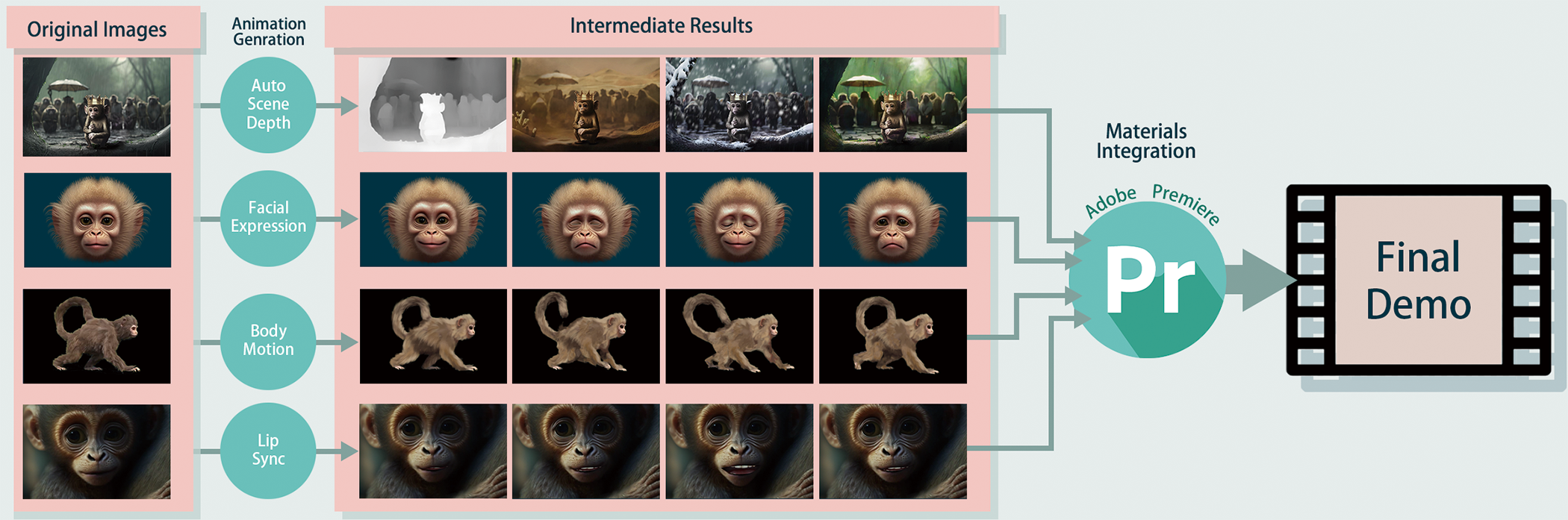}
    \caption{The overview of the animation production pipeline. We divide the animation section into four categories and apply different AIGC technologies integrations to achieve the animated effects. Then the relevant dynamic results are assembled using Adobe software to synthesize the whole clip. We strongly suggest visiting our website for better visualization: \href{https://lsgm-demo.github.io/Leveraging-recent-advances-of-foundation-models-for-story-telling/}{https://lsgm-demo.github.io/Leveraging-recent-advances-of-foundation-models-for-story-telling/}  (\textcopyright The author of the paper.)}
    \label{fig:figure5}
    \Description{The figure demonstrates the animation production pipeline. Four original images including a background scene, a monkey head, a monkey body, and a monkey face are individually sent into animation generation modules and produce multiple animation results. The animation results are then combined with Adobe Premiere. The final output is the demo video.}
\end{figure*}

To enhance the visual appeal and immerse the audience in the narrative, dynamic scenes were created by VoluMax AI and VoluMax Landscape in conjunction with Photoshop’s Neural Filters and their depth map engine, as shown in Figure~\ref{fig:figure5}. Interestingly, ControlNet’s depth detection algorithm ~\cite{zhang2023adding} was integrated into the scene animation production, enabling seamless weather and environmental transformations within the same scene.

Furthermore, character animations were produced through various technical combinations. We utilized Crazytalk’s facial animation function, which involved facial landmark detection, tracking, and expression mapping, and the First Order Motion Model (FOMM) to create lifelike facial expressions when the characters were not speaking, which added depth and emotional realism, making the characters relatable. To ensure seamless transitions between natural facial expressions in close-up shots, we integrated temporal smoothing and curve fitting features found in RE: Flex, further enhancing the overall visual experience.

However, the facial expression changes in the demo are insufficient and lack of real emotional expression. These limitations primarily stem from challenges in technology, data, and computing resources. Emotional expression is intricate and demands precise capturing of subtle facial expressions and body language. Current deep learning models are not yet sensitive enough to these minor emotional nuances. Furthermore, AIGC lacks the ability to recognize individual differences, making it difficult to generate unique vivid facial animations based on the characteristics and personalities of different characters.

Simultaneously, we used an automatic lip-syncing algorithm to synchronize the characters’ mouth movements with the lines generated by AI for realistic speech portrayal. Also, to efficiently animate characters in full body motion, texture synthesis and style transfer algorithms from Ebsynth were employed to build the animation out of certain static frames. Using this technique, the characters’ movements can be visually appealing and stylistically consistent throughout the production.

\begin{figure*}[h]
    \centering
    \includegraphics[width=1.0\textwidth]{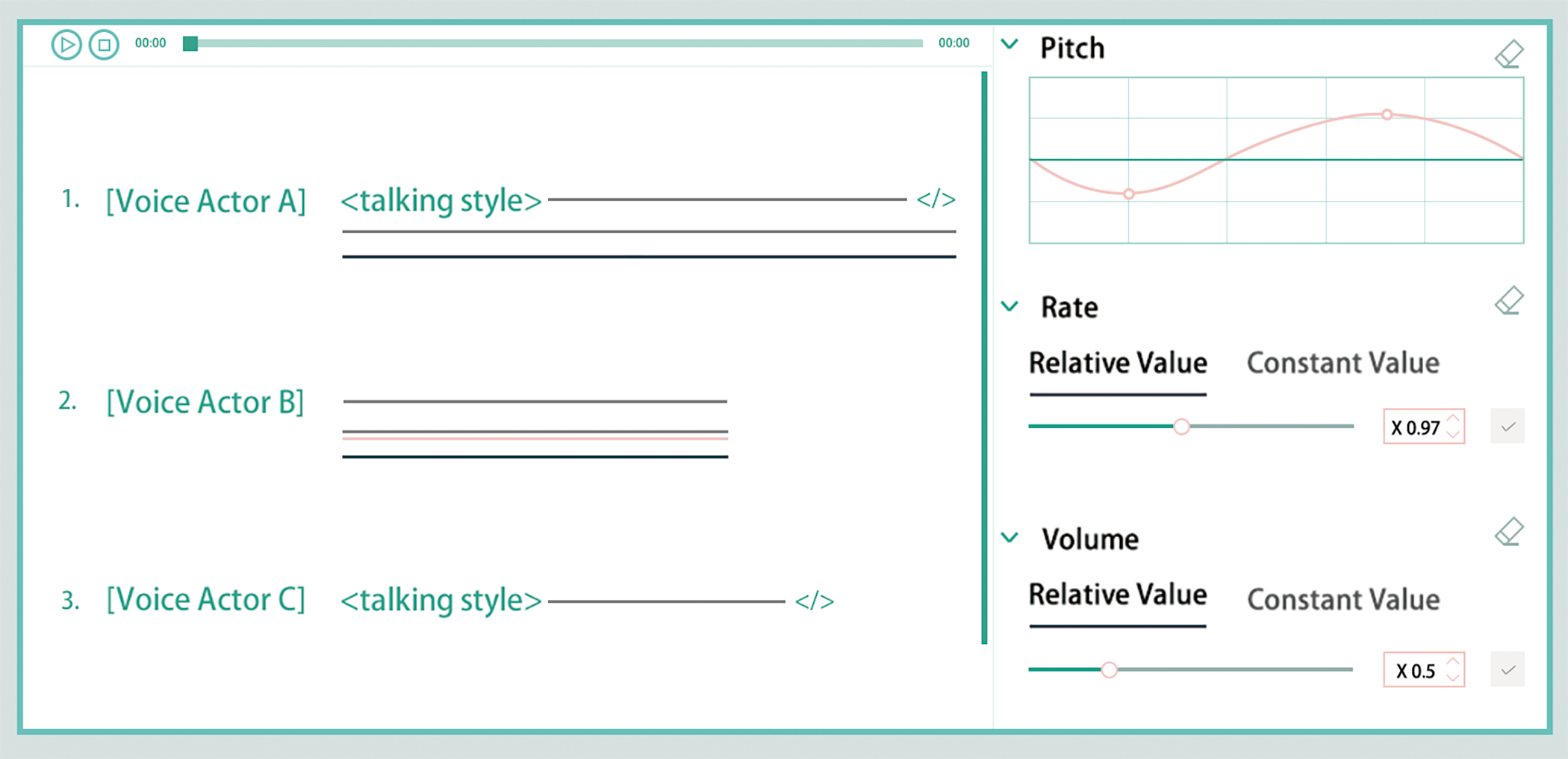}
    \caption{The schematic demonstration of customization function in TTS tool. The features of the TTS tool allow users to customize their AI-dubbed texts, including choosing voice actors and talking style, setting the pitch curve to match the ideal voice tone, and adjusting the rate and volume of each dubbing sentence (\textcopyright The author of the paper.)}
    \label{fig:figure6}
    \Description{The figure illustrates the audio editing dashboard. The scripts can be added for each actor's voice and marked with talking style variables. There are additional charts for editing the pitch conditions, rate parameters, and volume intensities.}
\end{figure*}

When comparing the animation effects achieved by current AIGC technologies with those produced through traditional processes, it becomes evident that AIGC is excellent at coordinating the dynamics of the characters when they are speaking, while its animation effect derived from AIGC-generated images is not competitive, particularly in terms of dynamic expression and three-dimensional motion. This observation highlights the considerable scope for improvement in authenticity, coherence, and personalization in the future development of AIGC.

\subsection{Audio}
TTS tools like Azure, illustrated in Figure~\ref{fig:figure6}, enable effective voice-over generation and voice acting for digital storytelling~\cite{tan2021survey}. Its extensive customization options, such as adjusting speaking rate, pitch, and volume~\cite{tan2021survey}, create immersive voice narration for allegory scripts and unique character voices, enhancing the emotional impact and audience engagement.

We also utilized Tianyin and SOUNDRAW to generate two AI-produced background music pieces that contain customized music length, tempo, composition, and instruments. These customized factors helped create the atmosphere and mood, build tension, create a sense of wonder, and amplify the storytelling experience.

Moreover, sound effects play a crucial role in digital storytelling by adding depth and realism to the narrative. Thus, we selected copyright-free sound effects from Freesound and processed all audio materials, such as reducing noise and adding reverb with Adobe Audition, to ensure the high quality of our audio.

Meanwhile, We endeavored to employ AI platforms such as MyEdit, Plugger.AI, and jsfxr, for the generation of sound effects; however, these attempts proved unsuccessful. This failure stemmed from the inherent requirement for sound effects in practical applications to exhibit a heightened level of contextual accuracy and emotional expression. Consequently, the efficacy of these sound effects necessitated a greater degree of verisimilitude to ensure that the audience could immerse themselves more completely in the auditory experience.

Furthermore, sound effects are tasked with capturing a diverse range of auditory elements that must be synchronized with the corresponding visual actions to achieve coherence and authenticity in the audio-visual narrative. In contrast, music production typically operates within a more abstract and emotionally driven framework, and it does not require precise synchronization with every visual action.

Consequently, AI-generated music and dubbing have undoubtedly achieved impressive results in practice, whereas sound effect production demands a higher degree of manual intervention and meticulous adjustment to satisfy the exacting requirements of the audio-visual experience. While the technology offers convenience and efficiency, it raises questions about authenticity and human creativity. It is essential to critically evaluate the balance between automation and artistic expression, ensuring that AI remains a tool rather than a replacement for human ingenuity.

\section{Voices}
We conducted insightful discussions with experts and artists from diverse creative industries to collect their feedback about the influences of applying AIGC within the digital storytelling arena. On the premise of introducing our experimental concept and prototype to the interviewees and acquainting them with AIGC’s current capabilities and performance, our discussions encompassed the interviewees’ perspectives, such as aesthetic connotations, substitutability, transformational development, and original creativity, on the current state of AIGC and its impact within their respective industries.

\subsection{Willingness}
After explaining our research concept and displaying the prototype, a critical examination of the interviewees’ responses reveals an overwhelmingly positive reception towards incorporating AIGC into art creation. A game industry expert highlighted the rapid development of AIGC and its impact on the art industry, he stated,

\begin{quote}
    \emph{AIGC benefits artistic creation and communication, and its rapid growth empowers emerging industry sectors.}
\end{quote}

They encouraged artists to embrace this trend for enhanced artistic expression. Henriikka and Matti~\cite{vartiainen2023using} mentioned that although AI's high automation can lead to conflicts in the actualization of creativity due to the lack of complementary resources, it is precisely these conflicts that serve as catalysts for innovation. However, it is important to critically analyze the extent to which AI-driven creations can capture the essence of human creativity and emotional depth.

An interviewee from the Human-Computer Interaction (HCI) domain saw AIGC as an opportunity to generate perspectives beyond human imagination:

\begin{quote}
    \emph{During human evolution, thinking tends to become fixed, but AI might inspire humans in the form of machine language.}
\end{quote}

This raises questions about the nature of creativity and whether AI can offer unique aesthetic values. It also prompts us to evaluate the role of AI in rediscovering neglected artistic elements and the risk of homogenization or dilution of artistic diversity.

An AI educator emphasized the value of current AI technology as a tool for enhancing artists’ efficiency in the creative process:

\begin{quote}
    \emph{From an artistic design standpoint, AIGC can produce various concept diagrams based on artists' input prompts, making early-stage concept creation more efficient and helping artists find inspiration or make selections.}
\end{quote}

The above viewpoints are consistent with Hong et al.~\cite{hong2019artificial}, which found that integrating AI into art creation does not negatively affect the artistic process or alter how viewers perceive the artistic value of the resulting works. However, it is crucial to assess the potential consequences of excessive reliance on AIGC. Over-reliance might hinder the development of essential artistic skills and discourage artists from exploring their own creativity, leading to dependence on preprogrammed algorithms for artistic expression.

\subsection{Replacement}
The interviewees were all questioned about their concerns regarding the potential replacement of human artists by AI. Without exception, they firmly believe that the current forms of AIGC are incapable of completely supplanting humans in the art industry. Marian et al.~\cite{mazzone2019art} substantiated this viewpoint, asserting that AI and artists don't compete but collaborate in artistic creation, dispelling the notion that AI could replace artists. One interviewee who specializes in HCI pointed out the fundamental flaw in AI’s capacity to comprehend human values and emotions:

\begin{quote}
    \emph{AI cannot replace essential work in the field of art. The most valuable aspect for artists is their creativity, which relies on human values and emotions. AI can only supplant rudimentary labor-intensive tasks and is inherently deficient in comprehending human values and emotions.}
\end{quote}

Another respondent working in the animation field argues that current AI technology falls short compared to humans when it comes to crucial creative tasks such as designing camera language as she stated,

\begin{quote}
    \emph{While using AIGC can save resources and time, replacing directors and producers is currently impossible because AI cannot replicate the aesthetic sensibilities and creative insights derived from human filmmakers.}
\end{quote}

An educator who teaches AI underscores the indispensable role played by human artists in selecting and curating art elements with their “unutterable power”, despite the remarkable visual generation by AIGC. In the context of the production process, Hong W also acknowledges this perspective by positing that AI is a product of data training, and its innovation is largely contingent upon the updating and iteration of algorithmic models. In terms of creativity, there remains a notable disparity between AI and human artists.

These perspectives are mirrored in our own work, as the artwork produced through these AI-driven processes often lacks human reasoning and spirit. Consequently, it becomes evident that the art industry still heavily relies on the unique contributions of human artists.

Artists typically express themselves through their unique perspectives, emotions, and creativity. While AIGC can generate content, they lack the depth of emotion, experience, and human thought, making it challenging to replace the originality of artists. The essence of artists lies in their creativity and uniqueness, whereas AIGC serves more as a tool to enhance or expand upon an artist's creativity.

\subsection{New Industry Format}
Experts in the game, animation, and HCI industries agree that AI has shown potential in optimizing art production and simplifying digital storytelling. They also believe that AI’s rapid development will lead to industry shifts, potentially creating more convenient art workflows and a new industry form during society’s digital transformation, even redistributing social values. This general sentiment is echoed by many theorists and practitioners, such as Chen et al.~\cite{chen2023challenges}, Gao et al.~\cite{gao2023aigc}, and Li et al.~\cite{li2023aigc}. However, AI will not completely replace any of the current creative industries because it is in a stage of rapid development, during which the new format is still unclear, and the industry regulations have not yet been standardized and improved to address AI specifically.

\subsection{The controls of AI}
After we presented our experiment result to our interviewees, over half of the interviewees expressed apprehension about the potential consequences of AIGC’s increasing dominance, raising valid concerns about the urgency to establish appropriate standards. A professor in the AI industry took photo contests as an example:

\begin{quote}
    \emph{Should AI-generated photos be considered on par with real-life scenario shots in a photo contest in terms of artistic merit? Is it necessary to assess these two types of creations separately? Furthermore, who holds the copyright for AI-generated photos?}
\end{quote}

It is evident that artists harbor a genuine fear of relinquishing control over their creative process and becoming mere conduits for AI-generated productions. This anxiety may stem from the realization that AIGC’s influence on art can be counterproductive. Furthermore, evaluating AI-assisted artwork presents a complex challenge due to inconsistent quality and the necessity for guidance on aesthetic enhancement.

Meanwhile, AIGC applications are increasingly impacting various industries through text, images, and audio generation, raising concerns about privacy, security, and copyright issues. To address these challenges, Chen et al.~\cite{chen2023challenges1} has previously proposed solutions such as integrating advanced technologies like blockchain and privacy computing to enhance user privacy and security and strengthening and enforcing relevant laws to tackle AIGC-related copyright problems.

Experts believe that the development of comprehensive standards and the prioritization of human decision-making are crucial steps to effectively integrate AIGC into the artistic realm. These measures are imperative to mitigate the potential risks and ensure that AI serves as a valuable tool rather than overshadowing human creativity.

\subsection{Summary}
In an era of inevitable technological revolution, the development of traditional art and AIGC occurs in parallel, a noncompetitive relationship that perpetually adheres to the aesthetic consciousness of creators. Technical professionals, including AIGC practitioners and other engineers, are responsible for developing and maintaining artificial intelligence systems. Their role is to ensure the smooth application of technology in the realm of art while also understanding the essence and needs of art to better support artists and creators. Technology serves as the foundation of AIGC, and its data originates from the market, so its developmental principles should be centered around creativity, rather than utilizing relatively singular visual representations that overshadow the diversity of artistic expressions.

Looking ahead, a human-level intelligent tool like AIGC helps drive the fusion of art creation and technological innovation, and promotes the democratization of art, and makes artistic expression freer and fairer while urging the public to have a deeper reflection on the nature of art. Thus, when using this inevitably derived new industry form, AIGC practitioners should continuously enhance their aesthetic sense and refine creative ideas from their very roots.

Under the circumstance that this dynamic field constantly introduces new industry forms and technologies, requiring practitioners to remain flexible and incorporate novel tools and methods into their work, cultivating a strong aesthetic sense is significant. Practitioners should start the creative process with meticulous ideation and concept development, rather than rushing into production, ensuring that creative ideas are nurtured and refined from their inception, thus setting a robust foundation for the generation of high-quality AIGC content.

To achieve this, AIGC practitioners should embrace a multifaceted approach. The specific suggestions involve staying updated with design trends and studying design history to draw inspiration from the masters. They should actively seek out constructive criticism, collaborate with diverse teams, and understand user psychology to create designs that resonate deeply. This path is the rational development essence for future AIGC professionals instead of blindly following the highly efficient expressive capabilities bestowed by technological advancements.

\section{Conclusion}
This study represents a guiding attempt in the transition of the traditional art industry toward incorporating AI as part of the creative process. The focal project of this study allows extensive investigation of the boundaries of mainly text-driven AIGC creative tools, and effective possibilities contained in various AIGC software and platforms, affirming their generation quality while also highlighting their limitations. 

Furthermore, this research serves as a guide, refinement, and catalyst for the current stage of incomplete AIGC creative standards and industry norms. The experimental portion contributes to a more compsssrehensive understanding of AIGC's current capabilities, while the insights that have been gathered through communications with abundant creative industry insiders section plays a beneficial role in directing the future growth of AIGC in a positive manner.

\noindent
\textbf{Acknowledgement.} We would like to express our gratitude to Shikai Li, Kwan-Yee Lin, and Huiwen Luo for their constructive suggestions and support throughout the research process. Additionally, we would like to acknowledge the Shanghai AI Laboratory and OpenXDLab that provided resources and facilities that were crucial to the completion of this work.
\bibliographystyle{ACM-Reference-Format}
\bibliography{sample-base}

\end{document}